# On the Normalization and Visualization of Author Co-Citation Data: Salton's Cosine *versus* the Jaccard Index

Loet Leydesdorff
Amsterdam School of Communications Research (ASCoR)
Kloveniersburgwal 48, 1012 CX  Amsterdam, The Netherlands
loet@leydesdorff.net ; http://www.leydesdorff.net

## Abstract

The debate about which similarity measure one should use for the normalization in the case of Author Co-citation Analysis (ACA) is further complicated when one distinguishes between the symmetrical co-citation—or, more generally, co-occurrence—matrix and the underlying asymmetrical citation—occurrence—matrix. In the Web environment, the approach of retrieving original citation data is often not feasible. In that case, one should use the Jaccard index, but preferentially after adding the number of total citations (occurrences) on the main diagonal. Unlike Salton's cosine and the Pearson correlation, the Jaccard index abstracts from the shape of the distributions and focuses only on the intersection and the sum of the two sets. Since the correlations in the co-occurrence matrix may partially be spurious, this property of the Jaccard index can be considered as an advantage in this case.

## 1. Introduction

Ahlgren *et al*. (2003) argued that one should consider using Salton's cosine instead of the Pearson correlation coefficient as a similarity measure in author co-citation analysis and showed the effects of this change on the basis of a dataset provided in Table 7 (at p. 555) of their paper. This has led to discussions in previous issues of this journal about the pros and cons of using the Pearson correlation or other measures (Ahlgren *et al*., 2004; Bensman, 2004; White, 2003, 2004; Leydesdorff, 2005). Leydesdorff and Vaughan (2006) used the same dataset for showing why one should use the (asymmetrical) citation instead of the (symmetrical) co-citation matrix as the basis for the normalization. They argued that not only the value, but also the sign of the correlation may change between two cited authors when using the Pearson correlation in the symmetrical *versus* the asymmetrical case. For example in the dataset under study, Ahlgren et al. (2003, at p. 556) found a correlation of $r = + 0.74$ between "Schubert" and "Van Raan," while Leydesdorff & Vaughan (at p. 1620) report $r = – 0.131$  ($p < 0.05$) using the underlying citation matrix.

One can download a set of documents in which the authors under investigation are potentially (co-)cited in the library environment, but this approach of retrieving original citation data and then using Pearson's $r$ or Salton's cosine to construct a similarity matrix is often not feasible in the web environment. In this environment, the researcher may only have the index available and searches the database with a Boolean AND in order to construct a co-citation or, more generally, a co-occurrence matrix without first generating an occurrence matrix. Should one in such cases also normalize using the cosine or the Pearson correlation coefficient or, perhaps, use still another measure?

I shall argue that in this case, one may prefer to use the Jaccard index (Jaccard, 1901). The Jaccard index was elaborated by Tanimoto (1957) for the non-binary case. Thus, one can distinguish between using the Jaccard index for the normalization of the binary citation matrix and the Tanimoto index in the case of the non-binary co-citation matrix. The results will be compared with using Salton's cosine (Salton & McGill, 1983), the Pearson correlation, and the probabilistic activity index (Zitt *et al.*, 2000) in the case of both the symmetrical co-citation and the asymmetrical citation matrix.

The argument is illustrated with an analysis using the same data as Ahlgren *et al*. (2003). This dataset (provided in Table 1) is extremely structured: it contains exclusively positive correlations within both groups and negative correlations between the two groups. The two groups are thus completely separated in terms of the Pearson correlation coefficients. However, there are relations between individual authors in the two groups. An optimal representation should reflect both this complete separation in terms of correlations at the level of the set and the weak overlap generated by individual relations (Waltman & Van Eck, forthcoming; Leydesdorff, 2005). (A visualization of the co-citation matrix before normalization is provided as Figure 13 by Leydesdorff & Vaughan (2006, at p. 1625).)

In summary, two problems have to be distinguished: the problem of normalization and the type of matrix to be normalized. In principle, one can normalize both symmetrical and asymmetrical matrices with the various measures. Ahlgren *et al*. (2003) provided arguments for using the cosine instead of the Pearson correlation coefficient, particularly if one aims at visualization of the structure like in the case of social network analysis or MDS. Bensman (2004) provided arguments why one might nevertheless prefer the Pearson correlation coefficient when the purpose of the study is a statistical (e.g., multivariate) analysis. The advantage of the cosine of being not a statistics, but a similarity measure then disappears. Formally, these two measures are equivalent with the exception that Pearson normalizes for the arithmetic mean, while the cosine doesn't use this mean as a parameter (Jones & Furnas, 1997). The cosine normalizes for the geometrical mean. The question remains which normalization one should use when one has only co-occurrence data available.

## 2. The Jaccard index

In his original paper introducing co-citation analysis, Small (1973, at p. 269) suggested the following solution to the normalization problem in footnote 6:

> We can also give a more formal definition of co-citation in terms of set theory notation. If A is the set of papers which cites document *a* and B is the set which cites *b*, then $A \cap B$, that is $n(A \cap B)$, is the co-citation frequency. The relative co-citation frequency could be defined as $n(A \cap B) \div n(A \cup B)$.

This proposal for the normalization corresponds with using the Jaccard index or its extension (for the non-binary case) into the Tanimoto index. The index is defined for a pair of vectors, $\mathbf{X}_m$ and $\mathbf{X}_n$, as the size of the *intersection* divided by the size of the union of the sample sets or in numerical terms:

$$S_{mn} = \frac{X_{mn}}{X_{mm} + X_{nn} - X_{mn}}$$

where $X_{ij} = \mathbf{X}_i\mathbf{X}_j$. The value of $S_{mn}$ ranges from 0 to 1 (Lipkes, 1999; cf. Salton & McGill, 1983, at pp. 203f.).

In a number of studies (e.g., Egghe & Rousseau, 1990; Glänzel, 2001; Hamers *et al.*, 1989; Leydesdorff & Zaal, 1988; Luukkonen *et al.*, 1993; Michelet, 1988; Wagner & Leydesdorff, 2005), the Jaccard index and the cosine have systematically been compared for co-occurrence data, but this debate has remained inconclusive. Using co-authorship data, for example, Luukkonen *et al.* (1993, at p. 23) argued that "the Jaccard measure is preferable to Salton's measure since the latter underestimatess the collaboration of smaller countries with larger countries; […]." Wagner & Leydesdorff (2005, at p. 208) argued that "whereas the Jaccard index focuses on strong links in segments of the database the Salton Index organizes the relations geometrically so that they can be visualized as structural patterns of relations."

In many cases, one can expect the Jaccard and the cosine measures to be monotonic to each other (Schneider & Borlund, forthcoming). However, the cosine metric measures the similarity between two vectors (by using the angle between them), whereas the Jaccard index focuses only on the relative size of the intersection between the two sets when compared to their union. Furthermore, one can normalize differently using the margintotals in the asymmetrical occurrence *or* the symmetrical co-occurrence matrix. Luukkonen *et al.* (1993, at p. 18), for example, summed the *co-occurrences* in their set (of 30 countries) for obtaining the denominator, while Small's (1973) definition of a relative co-citation frequency suggests to use the sum of the total number of *occurrences* as the denominator. White & Griffith (1981, at p. 165) also proposed using "total citations" as values for the main diagonal, but these authors decided not to use this normalization for empirical reasons.

Table 1 illustrates the two options by providing the data for the set under study and adding the total number of citations as the main diagonal and the total number of co-citations as margintotals. For example, using the non-binary margintotals for Schubert and Van Raan, respectively, the Tanimoto index is 5 / (139 + 132 – 5) = 0.019, while the Jaccard index based on the binary citations is 5 / (60 + 50 – 5) = 0.048. Important is that the co-occurrence matrix itself no longer informs us about the number of cited documents. The co-occurrence matrix contains less information than the occurrence matrix.[1]

[1] Two symmetrical matrices can be derived from one asymmetrical matrix. Borgatti *et al.* (2002) formulate this (in the manual of UCINet) as follows: "Given an incidence matrix A where the rows represent actors and the columns events, then the matrix AA' gives the number of events in which actors simultaneously attended. Hence AA' (i,j) is the number of events attended by both actor i and actor j. The matrix A'A

However, the total number of citations can be added by the researcher on the main diagonal. One could also consider this value as the search result for the co-citation of "Schubert AND Schubert," etc.

Note that the value added on the main diagonal of the co-citation matrix corresponds to the margintotal of the asymmetrical matrix, that is, the total number of citations. Therefore, a normalization of the symmetrical matrix using these values on the main diagonal precisely corresponds with using the Jaccard normalization of the asymmetrical occurrence matrix. I shall from hereon distinguish between the two normalizations in terms of the symmetrical and the asymmetrical matrix, respectively. In the latter case, I use the values on the main diagonal and in the former the margintotals.

Recall that the Jaccard index does not take the shape of the distributions in account, but only normalizes the intersection of two sets with reference to the sum of the two sets. In other words, the cell values are independently evaluated in relation to margintotals and not in relation to other cells in the respective rows and columns of the matrix. This insensitivity to the shape of the distributions can be considered as both an advantage and a disadvantage. In the case of the asymmetrical matrix, the Jaccard index does not exploit the full information contained in the matrix. This can be considered a disadvantage. Both the cosine and the Pearson correlation matrix fully exploit this information. However, in case of the symmetrical matrix one has already lost the information about the underlying distributions in the asymmetrical matrix. Import of the margintotals from the asymmetrical matrix as a value on the main diagonal then adds to the information contained in the symmetrical matrix.

The Jaccard index has this focus on cell values instead of distributions in common with the probabilistic activity index (PAI) which is the preferred measure of Zitt *et al.* (2000). The PAI is the (traditional) ratio between observed and expected values in a contingency table based on probability calculus:

$$\begin{aligned} PAI &= p_{ij} / (p_i * p_j) \\ &= n_{ij} * \Sigma_i \Sigma_j n_{ij} / \Sigma_i n_{ij} * \Sigma_j n_{ij} \end{aligned}$$

Like the Jaccard and Tanimoto index this index can be applied on the lower triangles of symmetrical co-occurrence matrices while the Pearson coefficient and the cosine are based on full vectors and thus use the information contained in a symmetrical matrix twice (Hamers *et al.*, 1989).[2]

---

gives the number of events simultaneously attended by a pair of actors. Hence A'A(i,j) is the number of actors who attended both event i and event j."

[2] Leydesdorff (2005) discussed the advantages of using information measures for the precise calculation of distances using the same co-occurrence data. Information theory is also based on probability calculus (cf. Van Rijsbergen (1977). The information measure generates an asymmetrical matrix based on a symmetrical co-occurrence matrix because the distance from A to B can be different from the distance between B and A. The measure thus generates a directed graph, while the measures under discussion here generate undirected graphs. Directed graphs can be visualized using Waldo Tobler's Flow Mapper, available at http://www.csiss.org/clearinghouse/FlowMapper/.

## 3. Results

Table 2 provides the Spearman rank-order correlations among the lower triangles of the various similarity matrices under discussion. Spearman's *rho* is used instead of Pearson's *r* because objects in proximity matrices are based on dyadic relationships (Kenny et al., 2006); the assumption of independence required for parametric significance tests is violated (Schneider & Borlund, forthcoming).

The perfect rank-order correlation ($\rho = 1.00$; $p < 0.01$) between the cosine matrix derived from the asymmetrical citation matrix, and the Jaccard index based on this same matrix supports the *analytical* conclusions above about the expected monotonicity between these two measures (Schneider & Borlund, forthcoming). However, there are some differences in the values which matter for the visualization. Figures 1 and 2 provide visualizations using these two matrices of similarity coefficients, respectively.

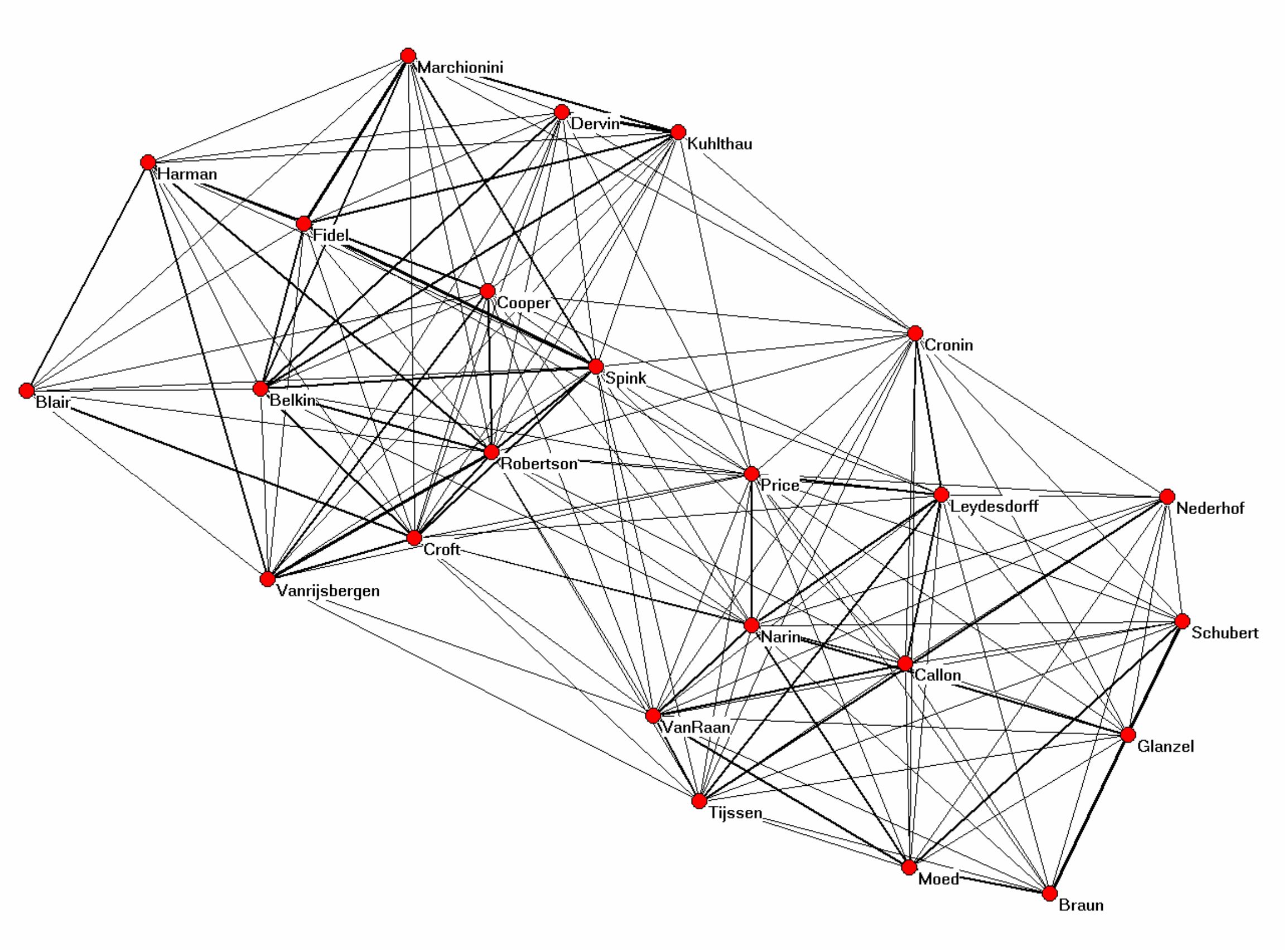

**Figure 1**: Cosine normalized representation of the asymmetrical citation matrix (Pajek;[3] Kamada & Kawai, 1989).

[3] Pajek is a software package for social network analysis and visualization which is freely available for academic usage at http://vlado.fmf.uni-lj.si/pub/networks/pajek/ .

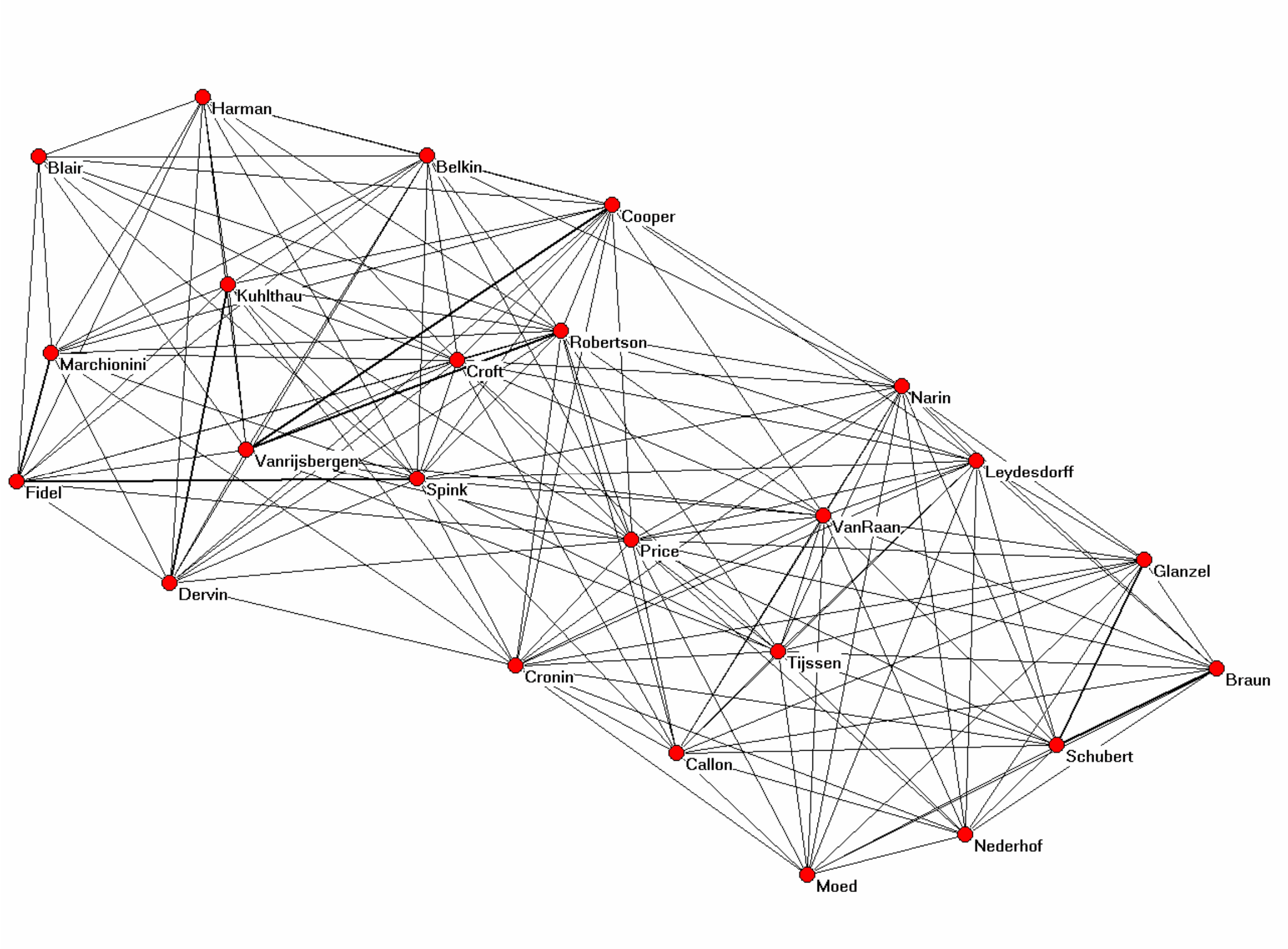

**Figure 2**: Jaccard-index-based representation of the co-citation matrix using total citations for the normalization (Pajek;[3] Kamada & Kawai, 1989))

The cosine remains the best measure for the visualization of the vector space because this measure is defined in geometrical terms. Although the Spearman correlation of the cosine-normalized matrix with the Jaccard index of this same matrix is unity, Figure 2 does not provide the fine-structure *within* the clusters to the same extent as Figure 1. Figure 3 shows that the Jaccard index covers a smaller range than the cosine (Hamers *et al.*, 1989). The smaller variance (0.08 versus 0.21 for the cosine-based matrix) may further limit the dissolvent capacity of the measure in visualizations.

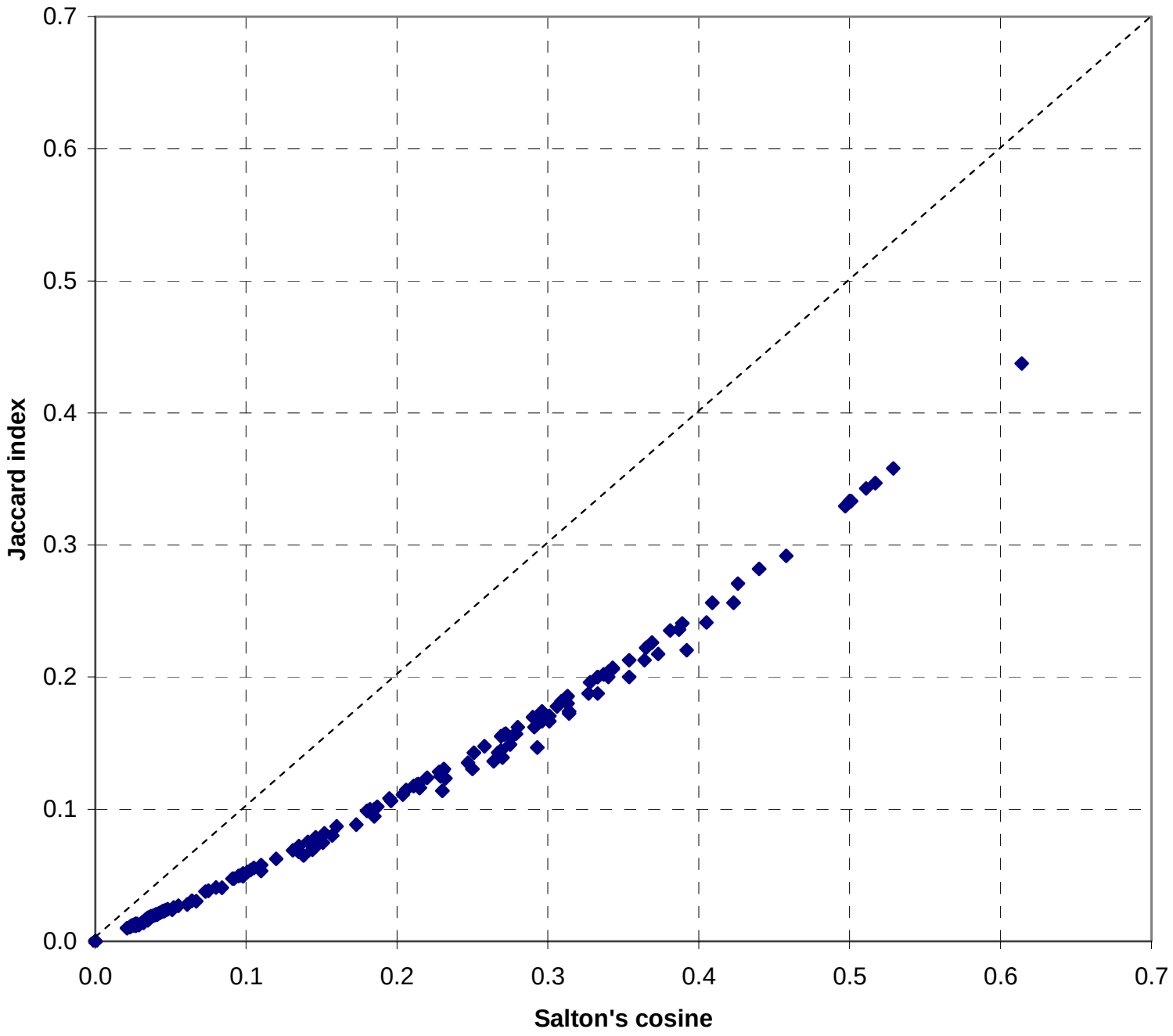

**Figure 3**: Relation between the Jaccard Index and Salton's cosine in the case of the asymmetrical citation matrix ($N = (24 * 23) / 2 = 276$).

In both cases, the analyst can emphasize the separation between the two groups by introducing a threshold. In the case of the Jaccard index the amount of detail in the relations between the two groups is then lower than in the case of the cosine-normalized matrix. For example, only the two co-citation relations between "Tijssen" and "Croft" pass a 0.05-threshold for the Jaccard index because both these authors have relatively low values on the main diagonal and therefore in the denominator of the equation, while several other co-citation relations (e.g., the relative intermediate positions of "Price" and "Van Raan") remain visible in the case of the cosine normalization and a cosine equal to or larger than 0.05.

The rank-order correlations of both these lower triangles with the Tanimoto index of the *symmetrical* matrix are also near unity ($\rho = 0.998$). All correlations with the probabilistic affinity index are slightly lower ($\rho < 0.99$). The correlations between using the Pearson correlation or the cosine on the asymmetrical and symmetrical matrices, respectively, are below 0.90. Despite the relatively small differences among the lower triangles, the visualizations are different.

In summary, the cosine-normalized asymmetrical occurrence matrix provides us with the best visualization of the underlying structure. When one is not able to generate an occurrence matrix, the Jaccard index using the values of the total number of citations on the main diagonal for the normalization is the second best alternative. In Table 3, I report on the results of using the twelve scientometricians as a subset. The results confirm that the Jaccard index normalized this way leads to results very similar ($\rho > 0.99$; $p < 0.01$) to the cosine-normalized occurrence matrix.

## Conclusions

Leydesdorff & Vaughan (2006) have provided reasons for using the asymmetrical matrix underlying the co-occurrence matrix as a basis for multivariate analysis (e.g., MDS, clustering, factor analysis). For the purposes of visualization, the cosine is the preferred measure for the reasons given by Ahlgren *et al*. (2003), but for other statistical analyses one may prefer to normalize using the Pearson correlation coefficient (Bensman, 2004) or Euclidean distances (in the case of MDS).

If the only option is to generate a co-occurrence matrix, as is often the case in webometric research, the Jaccard index is the best basis for the normalization *because* this measure does not take the distributions along the respective vectors into account. Like the Jaccard index, the PAI focuses only on the strength of the co-occurrence relation. If available, however, the frequencies of the occurrences which are conventionally placed on the main diagonal can be expected to improve the normalization. In the empirical examples, this Jaccard Index was as good a measure as the cosine-normalized citation matrices. Remember that the research question was which similarity measure to use when the occurrence matrix cannot be retrieved.

Which of the two options for the normalization of the Jaccard index will be preferable in a given project depends on the research question and the availability of the data. However, one should be very cautious in using the symmetrical matrix as input to further statistical analysis because of the change of the size and potentially the sign of the correlation when multiplying the citation matrix with its transposed. Using the Jaccard index with the diagonal value based on the margintotals of the asymmetrical matrix circumvents this problem.

**Acknowledgement**
I am grateful to Liwen Vaughan and three anonymous referees for comments on previous drafts of this paper.

| | | | | | | | | | | | | | | | | | | | | | | | | | |
|---|---|---|---|---|---|---|---|---|---|---|---|---|---|---|---|---|---|---|---|---|---|---|---|---|---|
| Braun | **50** | 29 | 19 | 19 | 8 | 13 | 5 | 9 | 7 | 7 | 2 | 0 | 0 | 0 | 0 | 0 | 0 | 0 | 0 | 0 | 0 | 0 | 0 | 0 | **118** |
| Schubert | 29 | **60** | 30 | 18 | 10 | 20 | 5 | 5 | 5 | 14 | 2 | 1 | 0 | 0 | 0 | 0 | 0 | 0 | 0 | 0 | 0 | 0 | 0 | 0 | **139** |
| Glanzel | 19 | 30 | **53** | 16 | 10 | 22 | 9 | 14 | 9 | 11 | 5 | 3 | 0 | 0 | 0 | 0 | 0 | 0 | 0 | 0 | 0 | 0 | 0 | 0 | **148** |
| Moed | 19 | 18 | 16 | **55** | 11 | 20 | 5 | 17 | 14 | 12 | 6 | 4 | 0 | 0 | 0 | 0 | 0 | 0 | 0 | 0 | 0 | 0 | 0 | 0 | **142** |
| Nederhof | 8 | 10 | 10 | 11 | **31** | 12 | 8 | 13 | 7 | 4 | 4 | 2 | 0 | 0 | 0 | 0 | 0 | 0 | 0 | 0 | 0 | 0 | 0 | 0 | **89** |
| Narin | 13 | 20 | 22 | 20 | 12 | **64** | 11 | 20 | 21 | 20 | 11 | 9 | 0 | 0 | 1 | 1 | 0 | 0 | 1 | 1 | 0 | 0 | 0 | 0 | **183** |
| Tijssen | 5 | 5 | 9 | 5 | 8 | 11 | **22** | 13 | 10 | 5 | 6 | 1 | 0 | 1 | 2 | 1 | 0 | 0 | 0 | 1 | 0 | 0 | 0 | 0 | **83** |
| VanRaan | 9 | 5 | 14 | 17 | 13 | 20 | 13 | **50** | 13 | 12 | 11 | 6 | 0 | 1 | 2 | 1 | 0 | 0 | 0 | 1 | 0 | 0 | 0 | 0 | **138** |
| Leydesdorff | 7 | 5 | 9 | 14 | 7 | 21 | 10 | 13 | **46** | 18 | 14 | 9 | 1 | 0 | 1 | 1 | 0 | 0 | 0 | 2 | 0 | 0 | 0 | 0 | **132** |
| Price | 7 | 14 | 11 | 12 | 4 | 20 | 5 | 12 | 18 | **54** | 10 | 9 | 1 | 1 | 1 | 1 | 0 | 0 | 2 | 0 | 1 | 0 | 1 | 2 | **132** |
| Callon | 2 | 2 | 5 | 6 | 4 | 11 | 6 | 12 | 14 | 10 | **26** | 4 | 0 | 0 | 1 | 1 | 0 | 0 | 0 | 1 | 0 | 0 | 0 | 0 | **79** |
| Cronin | 0 | 1 | 3 | 4 | 2 | 9 | 1 | 6 | 9 | 9 | 4 | **24** | 1 | 0 | 0 | 0 | 0 | 0 | 0 | 1 | 0 | 1 | 1 | 1 | **53** |
| Cooper | 0 | 0 | 0 | 0 | 0 | 0 | 0 | 0 | 1 | 1 | 0 | 1 | **30** | 14 | 5 | 11 | 5 | 8 | 6 | 2 | 0 | 0 | 1 | 1 | **56** |
| Vanrijsbergen | 0 | 0 | 0 | 0 | 0 | 0 | 1 | 1 | 0 | 1 | 0 | 0 | 14 | **30** | 7 | 15 | 5 | 13 | 5 | 3 | 1 | 0 | 1 | 1 | **68** |
| Croft | 0 | 0 | 0 | 0 | 0 | 1 | 2 | 2 | 1 | 1 | 1 | 0 | 5 | 7 | **18** | 9 | 6 | 7 | 8 | 6 | 2 | 1 | 2 | 2 | **63** |
| Robertson | 0 | 0 | 0 | 0 | 0 | 1 | 1 | 1 | 1 | 1 | 1 | 1 | 11 | 15 | 9 | **36** | 7 | 12 | 11 | 10 | 8 | 5 | 4 | 4 | **103** |
| Blair | 0 | 0 | 0 | 0 | 0 | 0 | 0 | 0 | 0 | 0 | 0 | 0 | 5 | 5 | 6 | 7 | **18** | 9 | 4 | 2 | 2 | 2 | 0 | 0 | **42** |
| Harman | 0 | 0 | 0 | 0 | 0 | 0 | 0 | 0 | 0 | 0 | 0 | 0 | 8 | 13 | 7 | 12 | 9 | **31** | 9 | 5 | 5 | 3 | 1 | 1 | **73** |
| Belkin | 0 | 0 | 0 | 0 | 0 | 1 | 0 | 0 | 0 | 2 | 0 | 0 | 6 | 5 | 8 | 11 | 4 | 9 | **36** | 9 | 9 | 10 | 14 | 10 | **98** |
| Spink | 0 | 0 | 0 | 0 | 0 | 1 | 1 | 1 | 2 | 0 | 1 | 1 | 2 | 3 | 6 | 10 | 2 | 5 | 9 | **21** | 11 | 7 | 5 | 4 | **71** |
| Fidel | 0 | 0 | 0 | 0 | 0 | 0 | 0 | 0 | 0 | 1 | 0 | 0 | 0 | 1 | 2 | 8 | 2 | 5 | 9 | 11 | **23** | 11 | 9 | 6 | **65** |
| Marchionini | 0 | 0 | 0 | 0 | 0 | 0 | 0 | 0 | 0 | 0 | 0 | 1 | 0 | 0 | 1 | 5 | 2 | 3 | 10 | 7 | 11 | **24** | 10 | 5 | **55** |
| Kuhlthau | 0 | 0 | 0 | 0 | 0 | 0 | 0 | 0 | 0 | 1 | 0 | 1 | 1 | 1 | 2 | 4 | 0 | 1 | 14 | 5 | 9 | 10 | **26** | 14 | **63** |
| Dervin | 0 | 0 | 0 | 0 | 0 | 0 | 0 | 0 | 0 | 2 | 0 | 1 | 1 | 1 | 2 | 4 | 0 | 1 | 10 | 4 | 6 | 5 | 14 | **20** | **51** |
| | **118** | **139** | **148** | **142** | **89** | **183** | **83** | **139** | **132** | **132** | **78** | **54** | **56** | **68** | **63** | **102** | **42** | **73** | **98** | **71** | **65** | **55** | **63** | **51** | **2,244** |

**Table 1**: Author co-citation matrix of 24 information scientists used (Table 7 of Ahlgren *et al*., 2003, at p. 555; main diagonal values added).

| | | | Pearson asymm. | cosine asymm. | Jaccard asymm. | Pearson symm. | cosine symm. | Tanimoto symm. | PAI symm. |
|---|---|---|---|---|---|---|---|---|---|
| Spearman's rho | Pearson asymmetrical | Correlation Coefficient | 1.000 | .910(**) | .909(**) | .828(**) | .818(**) | .904(**) | .910(**) |
| | | Sig. (2-tailed) | . | .000 | .000 | .000 | .000 | .000 | .000 |
| | | N | 276 | 276 | 276 | 276 | 276 | 276 | 276 |
| | Cosine asymmetrical | Correlation Coefficient | .910(**) | 1.000 | **1.000**(**) | .834(**) | .857(**) | **.998**(**) | .983(**) |
| | | Sig. (2-tailed) | .000 | . | .000 | .000 | .000 | .000 | .000 |
| | | N | 276 | 276 | 276 | 276 | 276 | 276 | 276 |
| | Jaccard asymmetrical | Correlation Coefficient | .909(**) | **1.000**(**) | 1.000 | .834(**) | .856(**) | **.998**(**) | .983(**) |
| | | Sig. (2-tailed) | .000 | .000 | . | .000 | .000 | .000 | .000 |
| | | N | 276 | 276 | 276 | 276 | 276 | 276 | 276 |
| | Pearson symmetrical | Correlation Coefficient | .828(**) | .834(**) | .834(**) | 1.000 | .818(**) | .837(**) | .823(**) |
| | | Sig. (2-tailed) | .000 | .000 | .000 | . | .000 | .000 | .000 |
| | | N | 276 | 276 | 276 | 276 | 276 | 276 | 276 |
| | Cosine symmetrical | Correlation Coefficient | .818(**) | .857(**) | .856(**) | .818(**) | 1.000 | .856(**) | .848(**) |
| | | Sig. (2-tailed) | .000 | .000 | .000 | .000 | . | .000 | .000 |
| | | N | 276 | 276 | 276 | 276 | 276 | 276 | 276 |
| | Tanimoto symmetrical | Correlation Coefficient | .904(**) | **.998**(**) | **.998**(**) | .837(**) | .856(**) | 1.000 | .984(**) |
| | | Sig. (2-tailed) | .000 | .000 | .000 | .000 | .000 | . | .000 |
| | | N | 276 | 276 | 276 | 276 | 276 | 276 | 276 |
| | PAI symmetrical | Correlation Coefficient | .910(**) | .983(**) | .983(**) | .823(**) | .848(**) | .984(**) | 1.000 |
| | | Sig. (2-tailed) | .000 | .000 | .000 | .000 | .000 | .000 | . |
| | | N | 276 | 276 | 276 | 276 | 276 | 276 | 276 |

** Correlation is significant at the 0.01 level (2-tailed).

**Table 2**: Spearman correlations among the lower triangles of similarity matrices using different criteria, and both asymmetrical citation and symmetrical co-citation data for 24 authors in both scientometrics and information retrieval.

| | | | Pearson asymm. | cosine asymm. | Jaccard asymm. | Pearson symm. | cosine symm. | Tanimoto symm. | PAI symm. |
|---|---|---|---|---|---|---|---|---|---|
| Spearman's rho | Pearson asymmetrical | Correlation Coefficient | 1.000 | .862(**) | .838(**) | -.042 | .253(*) | .766(**) | .912(**) |
| | | Sig. (2-tailed) | . | .000 | .000 | .736 | .040 | .000 | .000 |
| | | N | 66 | 66 | 66 | 66 | 66 | 66 | 66 |
| | Cosine asymmetrical | Correlation Coefficient | .862(**) | 1.000 | **.995**(**) | -.268(*) | .114 | .966(**) | .857(**) |
| | | Sig. (2-tailed) | .000 | . | .000 | .029 | .360 | .000 | .000 |
| | | N | 66 | 66 | 66 | 66 | 66 | 66 | 66 |
| | Jaccard asymmetrical | Correlation Coefficient | .838(**) | **.995**(**) | 1.000 | -.273(*) | .109 | .974(**) | .842(**) |
| | | Sig. (2-tailed) | .000 | .000 | . | .027 | .382 | .000 | .000 |
| | | N | 66 | 66 | 66 | 66 | 66 | 66 | 66 |
| | Pearson symmetrical | Correlation Coefficient | -.042 | -.268(*) | -.273(*) | 1.000 | .682(**) | -.256(*) | -.005 |
| | | Sig. (2-tailed) | .736 | .029 | .027 | . | .000 | .038 | .966 |
| | | N | 66 | 66 | 66 | 66 | 66 | 66 | 66 |
| | Cosine symmetrical | Correlation Coefficient | .253(*) | .114 | .109 | .682(**) | 1.000 | .069 | .190 |
| | | Sig. (2-tailed) | .040 | .360 | .382 | .000 | . | .582 | .127 |
| | | N | 66 | 66 | 66 | 66 | 66 | 66 | 66 |
| | Tanimoto symmetrical | Correlation Coefficient | .766(**) | .966(**) | .974(**) | -.256(*) | .069 | 1.000 | .837(**) |
| | | Sig. (2-tailed) | .000 | .000 | .000 | .038 | .582 | . | .000 |
| | | N | 66 | 66 | 66 | 66 | 66 | 66 | 66 |
| | PAI symmetrical | Correlation Coefficient | .912(**) | .857(**) | .842(**) | -.005 | .190 | .837(**) | 1.000 |
| | | Sig. (2-tailed) | .000 | .000 | .000 | .966 | .127 | .000 | . |
| | | N | 66 | 66 | 66 | 66 | 66 | 66 | 66 |

** Correlation is significant at the 0.01 level (2-tailed).
* Correlation is significant at the 0.05 level (2-tailed).

**Table 3**: Spearman correlations among the lower triangles of similarity matrices using different criteria, and both asymmetrical citation and symmetrical co-citation data for the subgroup of twelve scientometricians.